\documentclass[a4paper,oneside,final,notitlepage,onecolumn,12pt]{article}
\usepackage{amsfonts}
\usepackage{epsf}
\usepackage{graphicx}% Include figure files
\usepackage{amssymb,eso-pic}
\usepackage{latexsym}
\usepackage{tabularx}
\usepackage{amsxtra}
\usepackage{hyperref}
\usepackage{t1enc}
\usepackage{amsmath}
\usepackage{ifpdf,epsfig,array,amsmath,amssymb}

\usepackage{psfrag}
\psfrag{na}{\footnotesize $n^a$}   
\psfrag{ta}{\footnotesize $\tau^a$}   
\psfrag{vna}{\footnotesize $\check{n}^a$}
\psfrag{vNa}{\footnotesize $\check{N}^a$}
\psfrag{scrW}{\footnotesize $\mycal{W}_{\rho_0}$}
\psfrag{t=t1}{\footnotesize $\Sigma_{\tau_1}$}
\psfrag{t=t2}{\footnotesize $\Sigma_{\tau_2}$}
\psfrag{r=all}{\footnotesize $\rho=const$}
\psfrag{hna}{\footnotesize $\hat{n}^a$}
\psfrag{tna}{\footnotesize $\tilde{n}^a$}
\psfrag{hab,Kab}{\footnotesize $h_{ab}, K_{ab}$}
\psfrag{thab,tKab}{\footnotesize $\tilde h_{ab}, \tilde K_{ab}$}
\psfrag{hhab,hKab}{\footnotesize $\hat h_{ab}, \hat K_{ab}$}
\psfrag{chab,cKab}{\footnotesize $\check{h}_{ab}, \check{K}_{ab}$}

\usepackage[normalem]{ulem}
\usepackage{color}
\definecolor{blue}{rgb}{0,0,1}
\definecolor{red}{rgb}{1,0,0}


\DeclareFontFamily{OT1}{rsfs}{} \DeclareFontShape{OT1}{rsfs}{m}{n}{
<-7> rsfs5 <7-10> rsfs7 <10-> rsfs10}{}
\DeclareMathAlphabet{\mycal}{OT1}{rsfs}{m}{n}

\def\sc{{\hskip 3.5pt {{}^{{}^{{}_{{}_{\bowtie}}}}} \kern -8.pt{}}}  
\def\SC{{\hskip 3.5pt {{}^{{}^{{}^{{}_{{}_{\bowtie}}}}}} \kern -10.5pt{}}}

\begin{document}

%%%%%%%%%%%%%%%%%%%%%%%%%%%%%%%%%%%%%%%%%%%%%%%%%%%%%%%%%%%%%%%%%%%%%%%%%%%%%%
\newtheorem{theorem}{Theorem}[section]
\newtheorem{lemma}{Lemma}[section]
\newtheorem{proposition}{Proposition}[section]
\newtheorem{corollary}{Corollary}[section]
\newtheorem{conjecture}{Conjecture}[section]
\newtheorem{example}{Example}[section]
\newtheorem{definition}{Definition}[section]
\newtheorem{remark}{Remark}[section]
\newtheorem{exercise}{Exercise}[section]
\newtheorem{axiom}{Axiom}[section]
%%%%%%%%%%%%%%%%%%%%%%%%%%%%%%%%%%%%%%%%%%%%%%%%%%%%%%%%%%%%%%%%%%%%%%%%%%%%%%
\renewcommand{\theequation}{\thesection.\arabic{equation}} 
% A fenti parancs atdefinialja az egyenleteket szamozo parancsot
%%%%%%%%%%%%%%%%%%%%%%%%%%%%%%%%%%%%%%%%%%%%%%%%%%%%%%%%%%%%%%%%%%%%%%%%%%%%%%

\author{{Istv\'an R\'acz}\,\thanks{ ~email: racz.istvan@wigner.mta.hu}  
\\ %EndAName 
{Wigner RCP}\\  {H-1121 Budapest, Konkoly Thege Mikl\'os \'ut 29-33. Hungary
}}
\title{Is the Bianchi identity always hyperbolic?} 
%\title{On the determinacy based on lower dimensional contributions} 
%\title{On the nature of geometric determinacy} 

\maketitle

\begin{abstract}
We consider $n+1$ dimensional smooth Riemannian and Lorentzian spaces satisfying Einstein's equations. The base manifold is assumed to be smoothly foliated by a one-parameter family of hypersurfaces. In both cases---likewise it is usually done in the Lorentzian case---Einstein's equations may be split into `Hamiltonian' and `momentum' constraints and a `reduced' set of field equations. It is shown that regardless whether the primary space is Riemannian or Lorentzian whenever the foliating hypersurfaces are Riemannian the `Hamiltonian' and `momentum' type expressions are subject to a subsidiary first order symmetric hyperbolic system. Since this subsidiary system is linear and homogeneous in the `Hamiltonian' and `momentum' type expressions the hyperbolicity of the system implies that in both cases the solutions to the `reduced' set of field equations are also solutions to the full set of equations provided that the constraints hold on one of the hypersurfaces foliating the base manifold. 
\end{abstract}
%\noindent{\it pacs}: {04.25.D-}
%\noindent{\it Keywords}: Einstein's equation, gauge choice, evolution, multipole, spectral methods

\date

%"The whole problem with the world is that fools and fanatics are always so certain of themselves, but wiser men so full of doubts" Bertrand Russell

%%%%%%%%%%%%%%%%%%%% INTRODUCTION %%%%%%%%%%%%%%%%%%%%%%%%%%%%%%%%%%%%%%
\section{Introduction}\label{introduction}
\setcounter{equation}{0}

Consider a pair $(M,g_{ab})$, where $M$ is an $(n+1)$-dimensional ($n\geq 2$) smooth, paracompact, connected, orientable manifold endowed with a smooth metric $g_{ab}$ with signature which is either Euclidean or Lorentzian.\,\footnote{All of our other conventions will be as in \cite{wald}.} 

\medskip

Throughout this paper the geometry will be at the focus of our main concern. In restricting the geometry we shall assume that Einstein's  equations
\begin{equation}\label{geom}
G_{ab}-\mycal{G}_{ab}=0\,,
\end{equation}
holds, where, for simplicity, the source term $\mycal{G}_{ab}$ is assumed to have vanishing divergence. Note that whenever we have matter fields satisfying their field equations with energy-momentum tensor $T_{ab}$ and with cosmological constant $\Lambda$ the source term 
\begin{equation}
\mycal{G}_{ab} = 8\pi\,T_{ab} - \Lambda\,g_{ab}
\end{equation}
suits to the above requirements. 

\medskip

Concerning the topology of $M$ we shall assume that the manifold $M$ is foliated by a one-parameter family of hypersurfaces, i.e.~$M\simeq\mathbb{R}\times\Sigma$, for some codimension one manifold $\Sigma$. In other words, then $M$ possesses the structure of a trivial principal fiber bundle with structure group $\mathbb{R}$.

\medskip

Note that this assumption is known to hold \cite{geroch} for globally hyperbolic spacetimes but we would like to emphasize that as the signature of the metric may not be Lorentzian or even if it was, in deriving our key results, we need not to assume global hyperbolicity of the pertinent spacetime. Our assumptions on topology of $M$ are known to be equivalent to the existence of a smooth function $\sigma:M\rightarrow \mathbb{R}$ with non-vanishing gradient $\nabla_a \sigma$ such that the $\sigma=const$ level surfaces $\Sigma_{\sigma}=\{\sigma\}\times \Sigma$ comprise the one-parameter foliation of $M$.  

\medskip

Having the above generic setup it is natural to perform a $1+n$ decomposition. In doing so first a conventional $1+n$ splitting of (\ref{geom}) will be done by generalizing conventional arguments (see, e.g.~Section 2.4 of \cite{helmut&alan}). This $1+n$ splitting can be performed on equal footing in both the Lorentzian and Riemannian cases yielding `Hamiltonian' and `momentum' type expressions, along with a reduced set of equations referred as `evolutionary system'. By using the evolutionary system, a subsidiary system for the constraint expressions is derived. A remarkable and unexpected property of this subsidiary system is that regardless whether the metric of the imbedding manifold is of Lorentzian or Euclidean signature---whenever the metric on the $\sigma=const$ level surfaces is Riemannian---it comprises a first order symmetric hyperbolic system that is linear and homogeneous in the constraint expressions. This guaranties then that the constraint expressions vanish identically throughout domains where 
solutions to the evolutionary system exist provided that they vanish on one of the level surfaces. These results  are presented in Sections \ref{1+n_decomp}. Some useful relations are given in Section \ref{expl} and in the Appendix.

\medskip

Having been a $1+n$ type decomposition performed it is natural to ask whether analogous type of simplifications of the reduced equations in a succeeding $1+(n-1)$ splitting could also exist. The answer of this question requires---besides some obvious additional restrictions on the topology of the base manifold---the identification of those conditions which guarantee that the covariant divergence of the new source term ${}^{{}^{(n)}}\hskip-1mm \mycal{G}_{ab}$ vanishes. The corresponding analysis is carried out in Section \ref{integrability}. The main conclusion here is that even though a formal $1+(n-1)$ splitting could be performed, in general, there is no room to acquire additional new simplifications. What can be done is nothing more than a redistribution of the simplifications associated with the primary splitting of the original field equations. 

\medskip

The paper is closed in Section \ref{final} by remarks on some of the implications of the derived new results.

%%%%%%%%%%%%%%%%%%%% 1+n decomposition %%%%%%%%%%%%%%%%%%%%%%%%%%%%%%%%%%%
\section{The $1+n$ decomposition}\label{1+n_decomp}
\setcounter{equation}{0}

This section is to show that a reduced set of the equations can be deduced from (\ref{geom}) such that, regardless whether the metric of the imbedding manifold is of Lorentzian or Euclidean signature, the solutions to this reduced system are also solutions to the full set (\ref{geom}) provided that the `constraints' hold on one of the $\sigma=const$ level surfaces. 

\medskip

We are proceeding by separating the `evolution' and `constraint' equations by adopting the strategy of the conventional $1+3$ decomposition applied in spacetimes with Lorentzian metric (see, e.g.~\cite{helmut&alan}). 
In doing so denote by $n^a$ the `unit norm' vector field that is normal to the $\sigma=const$ level surfaces. To allow the simultaneous investigation of both spaces with either Euclidean or Lorentzian signature and timelike or spacelike level surfaces the sign of the norm of $n^a$ will not be fixed, i.e.~it will be assumed that
\begin{equation}%\label{SymS2}
n^a n_a=\epsilon\,,
\end{equation}
where $\epsilon$ takes the value $-1$ or $+1$. 

\medskip

The induced metric $h_{ab}$ and the pertinent projection operator ${h^a}_b$ on the level surfaces of $\sigma:M\rightarrow \mathbb{R}$ are then given as
\begin{equation}%\label{SymS2}
h_{ab}=g_{ab}-\epsilon\, n_a n_b\,, \hskip3mm {\mathrm and} \hskip3mm 
{h^a}_b={g^a}_b -\epsilon\,  n^a n_b\,,
\end{equation}
respectively.

\medskip

Denote by $E_{ab}$ the left-hand-side of (\ref{geom}), i.e.
\begin{equation}\label{ein2}
E_{ab}=G_{ab}-\mycal{G}_{ab}\,,
\end{equation}
and define the `Hamiltonian' $E^{{}^{(\mathcal{H})}}$ and `momentum' $E^{{}^{(\mathcal{M})}}_b$ expressions as
\begin{equation} 
E^{{}^{(\mathcal{H})}}=n^e n^f E_{ef} \,, \hskip3mm {\rm and} \hskip3mm 
E^{{}^{(\mathcal{M})}}_b=n^e {h^f}_b  E_{ef} \,,
\end{equation}
respectively. Then, we have 
\begin{equation}\label{constr01}
E_{ab}= {h^e}_a {h^f}_b\, E_{ef}+\epsilon\,[n_a\,E^{{}^{(\mathcal{M})}}_b+n_b\,E^{{}^{(\mathcal{M})}}_a] +n_a n_b\, E^{{}^{(\mathcal{H})}} \,.
\end{equation}

\medskip

Choose now as our `evolutionary' system the combination 
\begin{equation}\label{evol} 
E^{{}^{(\mathcal{EVOL})}}_{ab}\hskip-.1cm{}={h^e}_a {h^f}_b\, E_{ef} - \kappa\, h_{ab} \,E^{{}^{(\mathcal{H})}}=0 \,,
\end{equation}
where $\kappa$ is some constant. Then, by combining (\ref{constr01}) and (\ref{evol}),  we get
\begin{equation}\label{constr1}
E_{ab}= \epsilon\,[n_a\,E^{{}^{(\mathcal{M})}}_b+n_b\,E^{{}^{(\mathcal{M})}}_a] +\left[\left(1-\epsilon\,\kappa\right)\,n_a n_b+\kappa\, g_{ab}\right] E^{{}^{(\mathcal{H})}} \,.
\end{equation}

Taking now the $\nabla^a$ divergence of (\ref{constr1}) and using our assumption concerning the vanishing of the covariant divergence $\nabla^a\mycal{G}_{ab}$, along with the twice contracted Bianchi identity we get
\begin{eqnarray}
&& \epsilon\, (\nabla^a n_a) E^{{}^{(\mathcal{M})}}_b + \epsilon\,(n^a \nabla_a E^{{}^{(\mathcal{M})}}_b)  + \epsilon\,(E^{{}^{(\mathcal{M})}}_a \nabla^a n_b ) + \epsilon\,n_b\,(\nabla^a E^{{}^{(\mathcal{M})}}_a ) \label{divConst} \\
&& \hskip-1cm (1-\epsilon\,\kappa)\,\left\{\left[ (\nabla^a n_a)\, n_b + (n^a\nabla_a n_b)\right] E^{{}^{(\mathcal{H})}} + \,n_b\,(n^a \nabla_a E^{{}^{(\mathcal{H})}})\right\} + \kappa\,\nabla_b E^{{}^{(\mathcal{H})}}=0\,. \nonumber 
\end{eqnarray}
The `parallel' and `orthogonal' parts of (\ref{divConst}) read then as
\begin{eqnarray}
\hskip-.5cm n^e\nabla_e E^{{}^{(\mathcal{H})}} + \epsilon\,h^{ef} D_e E^{{}^{(\mathcal{M})}}_f {}&\hskip-.3cm=&\hskip-.2cm{} (1-\epsilon)\,(n^e \nabla_e n^b) \,E^{{}^{(\mathcal{M})}}_b - \epsilon\,\left(1- \epsilon\,\kappa\right)\,\left(\nabla_e n^e\right) E^{{}^{(\mathcal{H})}}\,, \label{constr2} \\
\hskip-.5cm h^{af}n^e\nabla_e E^{{}^{(\mathcal{M})}}_f + \epsilon\,\kappa\, h^{af} D_f E^{{}^{(\mathcal{H})}} {}&\hskip-.3cm=&\hskip-.3cm{} -h^{af} E^{{}^{(\mathcal{M})}}_f \hskip-.1cm\left(\nabla^e n_e\right) - E^{{}^{(\mathcal{M})}}_e \hskip-.1cm\left(\nabla^e n_f\right) h^{fa} \label{constr3}\\
&& \phantom{-h^{af}\,E^{{}^{(\mathcal{M})}}_f \left({}_a n_e\right) } 
-  \epsilon\,\left(1-\epsilon\,\kappa\right)h^{af}\left(n^e\nabla_e n_f\right) E^{{}^{(\mathcal{H})}}\,, \nonumber 
\end{eqnarray}
where the relations $\epsilon^2=1$ and
\begin{equation}%\label{SymS2}
\nabla^a E^{{}^{(\mathcal{M})}}_a=D^a E^{{}^{(\mathcal{M})}}_a-\epsilon\,(n^a\nabla_a n^b)\,E^{{}^{(\mathcal{M})}}_b  \,
\end{equation}
have been used, and $D_a$ denotes the unique torsion free covariant derivative operator associated with $h_{ab}$.

\medskip

Although $(M,g_{ab})$ may not have anything to do with time evolution we shall refer to a vector field $\sigma^a$ on $M$ as an `evolution vector field' if the relation $\sigma^e\nabla_e\sigma=1$ holds. Notice that this condition guaranties that $\sigma^a$ nowhere vanishes nor becomes tangent to the $\sigma=const$ level surfaces. The unit normal $n^a$ to these level surfaces may always be decomposed as 
\begin{equation}%\label{SymS2}
n^a=\frac1N\,\left[ (\partial_\sigma)^a-N^a\right]\,,
\end{equation}
where $N$ and $N^a$ denotes the `laps' and `shift' of the `evolution' vector field $\sigma^a=(\partial_\sigma)^a$ defined as 
\begin{equation}%\label{SymS2}
N= \epsilon\,(\sigma^e n_e) \hskip0.5cm {\rm and} \hskip0.5cm N^a={h^{a}}_{e}\,\sigma^e\,,
\end{equation}
respectively.
Taking these relations into account, equations (\ref{constr2}) and (\ref{constr3})---when writing them out explicitly in some local coordinates $(\sigma,x^{1},\dots,x^{n})$ adopted to the vector field $\sigma^a$ and the foliation $\{ \Sigma_{\sigma} \}$---can be seen to take the form\,\footnote{The spatial indices of the pull backs of geometrical objects to the $\sigma=const$ slices yielded in the applied $1+n$ decomposition will be indicated by lowercase Latin indices from the second half and they will be assumed to take the values $1,\dots,n$.}
\begin{equation}\label{evol_ls}
\left\{ \left(
\begin{array}{cc}
\frac1N & 0  \\
 0 & \frac1N\, h^{ij}  
\end{array} 
\right)\,\partial_\sigma +
\left(
\begin{array}{cc}
 -\frac1N\, N^{k} & \epsilon\,h^{ik}  \\
\epsilon\,\kappa\, h^{jk} & -\frac1N\, N^{k}\,h^{ij} 
\end{array} 
\right)\,\partial_k
\right\}
\left(
\begin{array}{c}
 E^{{}^{(\mathcal{H})}}   \\
 E^{{}^{(\mathcal{M})}}_i  
\end{array} 
\right) = 
\left(
\begin{array}{c}
 \mycal{E}  \\
 \mycal{E}^j  
\end{array} 
\right)\,,
\end{equation}
where, in virtue of (\ref{constr2}) and (\ref{constr3}), $\mycal{E}$ and $\mycal{E}^j$ are linear and homogeneous expressions of $E^{{}^{(\mathcal{H})}}$ and $E^{{}^{(\mathcal{M})}}_i$. It follows immediately that the coefficient matrices of the partial derivatives are symmetric if $\kappa=1$ and, in addition, the coefficient of $\partial_\sigma$ is also positive definite provided that the induced metric $h^{ij}$ is positive definite. 

\medskip

Hereafter we shall assume that $\kappa=1$ and that $h^{ij}$ is positive definite. The latter occurs if the $\sigma$ level surfaces are spacelike allowing the signature of metric $g_{ab}$ to be either Lorentzian, with $\epsilon=-1$, or Euclidean, with $\epsilon=+1$, respectively. In these cases (\ref{evol_ls}) comprises a first order symmetric hyperbolic linear and homogeneous system
\begin{equation}\label{evol_ls2}
\mathcal A^\mu \,\partial_\mu v + \mathcal B\, v =0 
\end{equation}
for the vector valued variable $v=(E^{{}^{(\mathcal{H})}},E^{{}^{(\mathcal{M})}}_i)^T$. As these type of equations are guaranteed to have identically vanishing solution for vanishing initial data the `Hamiltonian' and `momentum' expressions will be guaranteed to vanish throughout the domain of existence of solutions to the evolutionary system  (\ref{evol}), with $\kappa=1$, provided they vanish on one of the slides of the foliation $\{ \Sigma_{\sigma} \}$. 

\medskip

By combining the above observations we have the following:

\begin{theorem}\label{1stsym}\vskip-0.2cm
Let $(M,g_{ab})$ as described in Section \ref{introduction} such that the metric induced on the $\sigma=const$ level surfaces is Riemannian. Then, regardless whether $g_{ab}$ is of Lorentzian or Euclidean signature, any solution to $E^{{}^{(\mathcal{EVOL})}}_{ab}\hskip-.1cm{}=0$, with $\kappa=1$, is also a solution to the full set (\ref{geom}) provided that $E^{{}^{(\mathcal{H})}}$ and $E^{{}^{(\mathcal{M})}}_a$ vanish on one of the level surfaces.
\end{theorem}

It is a remarkable property of (\ref{evol_ls}) that $\epsilon$ and $\kappa$ do not show up in the coefficient of $\partial_\sigma$, and once $\kappa=1$ is chosen all the coefficients ${\mathcal A}^\mu$ in (\ref{evol_ls2}) are guaranteed to be symmetric regardless of the value of $\epsilon$. 

\section{The explicit forms}\label{expl}
\setcounter{equation}{0}

In exploring some of the consequences of Theorem \ref{1stsym} we shall need the explicit forms of the constraint expressions and the evolutionary system. In spelling out them we shall refer to the extrinsic curvature $K_{ab}$ which is defined as 
\begin{equation}\label{extcurv}
K_{ab}= {h^{e}}_{a} \nabla_e n_b=\frac12\,\mycal{L}_n  h_{ab}\,,
\end{equation}
where $\mycal{L}_n$ stands for the Lie derivative with respect to $n^a$.

\medskip

The `Gauss' and `Codazzi' relations take the form 
\begin{eqnarray}%\label{SymS2}
{h^{e}}_{a}{h^{f}}_{b} {h^{k}}_{c} {h^{d}}_{j}  {R_{efk}}^j {}&\hskip-.3cm=&\hskip-.3cm{} {{}^{{}^{(n)}}\hskip-1mm R_{abc}}{}^d-\epsilon\,\left\{ 
K_{ac}{K^d}_b-K_{bc}{K^d}_a\right\}\,,\\
{h^{e}}_{a}{h^{f}}_{b} n^k {h^{d}}_{j}  {R_{efk}}^j {}&\hskip-.3cm=&\hskip-.3cm{} D_b {K^d}_a-D_a {K^d}_b\,,
\end{eqnarray}
where ${{}^{{}^{(n)}}\hskip-1mm R_{abc}}{}^d$ stands for the $n$-dimensional Riemann tensor associated with $h_{ab}$.

\medskip

The various projections of the full Ricci tensor---which can be derived either by contractions of the above two relations or that of the third non-trivial projection of the full Riemann tensor, $n^a{h^{f}}_{b} n^c {h^{d}}_{j}  {R_{efk}}^j$---read as
\begin{eqnarray}%\label{SymS2}
&& \hskip-1.6cm {h^{e}}_{a}{h^{f}}_{b} R_{ef} {}= {{}^{{}^{(n)}}\hskip-1mm R_{ab}} + \epsilon\,\left\{\phantom{\frac{2\,\epsilon}{N}}\hskip-.6cm -\mycal{L}_n  K_{ab}- K_{ab}{K^e}_e + 2 K_{ae}{K^e}_b  - \frac{\epsilon}{N}\,D_a D_b N
\right\},\label{expl_hhR}\\
&& \hskip-1.6cm {h^{e}}_{a} n^f R_{ef} = D_e {K^e}_a-D_a {K^e}_e, \label{expl_hnR} \\
&& \hskip-1.6cm n^e n^f R_{ef} = -\left\{ \mycal{L}_n  ({K^e}_{e}) + K_{ef}K^{ef} + \frac{\epsilon}{N}\,D^e D_e N 
\right\},\label{expl_nnR}
\end{eqnarray} 
where ${}^{{}^{(n)}}\hskip-1mm R_{ab}$ stand for the Ricci tensor associated with $h_{ab}$.

\medskip

Taking all the above relations into account we have
\begin{eqnarray} 
E^{{}^{(\mathcal{H})}} {}&\hskip-.4cm=&\hskip-.2cm{} n^e n^f E_{ef}=\frac12\,\left\{ -\epsilon\,{}^{{}^{(n)}}\hskip-1mm R + \left({K^{e}}_{e}\right)^2 - K_{ef} K^{ef} - 2\,\mathfrak{e} \right\}\,, \label{expl_eh}\\
E^{{}^{(\mathcal{M})}}_a {}&\hskip-.4cm=&\hskip-.2cm{} {h^e}_a n^f  E_{ef}=D_e {K^{e}}_{a} - D_a {K^{e}}_{e} - \epsilon\,\mathfrak{p}_{a}\,,\label{expl_em} \\
E^{{}^{(\mathcal{EVOL})}}_{ab}\hskip-.1cm{} {}&\hskip-.4cm=&\hskip-.2cm{} {}^{{}^{(n)}}\hskip-1mm R_{ab} +\epsilon\,\left\{ -\mycal{L}_n K_{ab} - ({K^{e}}_{e}) K_{ab} + 2\,K_{ae}{K^{e}}_{b} - \frac{\epsilon}{N}\,D_a D_b N \right\}- \left[\mathfrak{S}_{ab}-\mathfrak{e}\,{h}_{ab}\right] \nonumber\\ 
&& \hskip-1.7cm -\frac12\,{h}_{ab}\,\left\{\phantom{\frac{2\,\epsilon}{N}}\hskip-.5cm (1-\epsilon)\,{}^{{}^{(n)}}\hskip-1mm R -2\,\epsilon\,\mycal{L}_n ({K^e}_{e}) + (1-\epsilon)\,\left({K^{e}}_{e}\right)^2 - (1+\epsilon)\,K_{ef} K^{ef} - \frac{2}{N}\,D^e D_e N  \right\} \,,\nonumber\\ \label{evol_ev} 
\end{eqnarray}
where $\mathfrak{e}= n^e n^f\,\mycal{G}_{ef}$, $\mathfrak{p}_{a}=\epsilon\,{h^{e}}_{a} n^f\, \mycal{G}_{ef}$ and $\mathfrak{S}_{ab}={h^{e}}_{a}{h^{f}}_{b}\,\mycal{G}_{ef}$.

\medskip

For certain cases (in particular, whenever $\epsilon=-1$) it is rewarding to do some algebra by which it can be verified that 
\begin{equation}\label{new_evol}
E^{{}^{(\mathcal{EVOL})}}_{ab}\hskip-.1cm{} -\frac1{n-1}\,{h}_{ab} \left(E^{{}^{(\mathcal{EVOL})}}_{ef} h^{ef}\right)= \widetilde  E^{{}^{(\mathcal{EVOL})}}_{ab}\hskip-.1cm{}\,,
\end{equation}
where
\begin{equation}\label{new_evol_2}
\widetilde E^{{}^{(\mathcal{EVOL})}}_{ab}\hskip-.1cm{}={h^{e}}_{a}{h^{f}}_{b}\,\left[{R}_{ab}-\left(\mycal{G}_{ab}-\frac1{n-1}\,g_{ab}\,[\mycal{G}_{ef}\,g^{ef}] \right)\right]
+\frac{1+\epsilon}{n-1}\,{h}_{ab}\,E^{{}^{(\mathcal{H})}} \,.
\end{equation}
In virtue of the above relations we have 
\begin{lemma}
The evolutionary system (\ref{evol}) holds if and only if either 
\begin{itemize}
\item[(i)]  the right hand side of (\ref{evol_ev}), or 
\item[(ii)] that of (\ref{new_evol_2}) 
\end{itemize}
vanishes. 
\end{lemma}

The right hand side of (\ref{new_evol_2}) can also be written as
\begin{eqnarray} 
\widetilde E^{{}^{(\mathcal{EVOL})}}_{ab}\hskip-.1cm{} {}&\hskip-.4cm=&\hskip-.2cm{} {}^{{}^{(n)}}\hskip-1mm R_{ab} +\epsilon\,\left\{ -\mycal{L}_n K_{ab} - ({K^{e}}_{e}) K_{ab} + 2\,K_{ae}{K^{e}}_{b} - \frac{\epsilon}{N}\,D_a D_b N \right\} \label{evol_ev_2} \\
&& \hskip-2.cm -\left(\mathfrak{S}_{ab}-\frac1{n-1}\,h_{ab}[\mathfrak{S}_{ef}\,h^{ef}+\epsilon\,\mathfrak{e} ] \right)
+\frac{1+\epsilon}{2\,(n-1)}\,{h}_{ab}\left\{ -\epsilon{}^{{}^{(n)}}\hskip-1mm R + \left({K^{e}}_{e}\right)^2 - K_{ef} K^{ef} - 2\,\mathfrak{e} \right\}. \nonumber
\end{eqnarray}

\medskip

Note that by making use of the contractions $\mathfrak{e}$, $\mathfrak{p}_{a}$ and $\mathfrak{S}_{ab}$ our source term $\mycal{G}_{ab}$ can be decomposed as
\begin{equation}
\mycal{G}_{ab}= n_a n_b\,\mathfrak{e} + \left[n_a \,\mathfrak{p}_b  + n_b\,\mathfrak{p}_a\right]  + \mathfrak{S}_{ab}\,,
\end{equation}
while its divergence $\nabla^a\mycal{G}_{ab}$ take the form [see also (\ref{Div_P_ab})]
\begin{eqnarray}\label{cons_eq}
&& \hskip-0.8cm \nabla^a \mycal{G}_{ab} = \mathfrak{e} \,({K^e}_e)\,n_b+ ({K^e}_e) \,\mathfrak{p}_b + \mathfrak{p}_e {K^e}_b + n_b\,(D^e \mathfrak{p}_e) + D^e \mathfrak{S}_{eb} - \epsilon\,n_b\,(\mathfrak{S}_{ef}K^{ef})  \nonumber \\ 
&& \hskip-0.8cm \phantom{\nabla_e P_{ab} = } + \dot n_b\,\mathfrak{e} + n_b\,\mycal{L}_{n}\mathfrak{e} + \mycal{L}_{n} \mathfrak{p}_b - \mathfrak{p}_e {K^e}_b - 2\,\epsilon\,(\dot n^e\mathfrak{p}_e)\,n_b - \epsilon\,(\dot n^e\mathfrak{S}_{eb})\,, 
\end{eqnarray}
where
\begin{equation}\label{dotn}
\dot n_a:=n^e\nabla_e n_a=-\epsilon\,D_a \ln N\,. 
\end{equation}
Taking then the `parallel' and `orthogonal' parts of (\ref{cons_eq}), 
\begin{equation}\label{cons_law}
\nabla^a\mycal{G}_{ab}=0%\,.
\end{equation}
we get [see also (\ref{par_Div_P_ab}) and (\ref{ort_Div_P_ab})]
\begin{eqnarray} 
\mycal{L}_n\,\mathfrak{e} + D^e \mathfrak{p}_e +\left[ \,\mathfrak{e}\,({K^e}_e) -2\,\epsilon\,( \dot n^e\,\mathfrak{p}_e) - \epsilon\,K^{ae}\,\mathfrak{S}_{ae} \,\right] {}&\hskip-.4cm=&\hskip-.2cm{} 0\,, \label{cons_law21} \\ 
\mycal{L}_n\,\mathfrak{p}_b + D^a\mathfrak{S}_{ab} + \left[- \epsilon\,\mathfrak{S}_{ab}\,\dot n^a + ({K^e}_e)\,\mathfrak{p}_b+ \mathfrak{e}\,\dot n_b   \,\right] {}&\hskip-.4cm=&\hskip-.2cm{} 0\,. \label{cons_law22}
\end{eqnarray}

\medskip

Notice that in deriving (\ref{cons_law21}) and (\ref{cons_law22}) only the vanishing of the divergence $\nabla^a\mycal{G}_{ab}$ has been used.\,\footnote{Relations analogous to (\ref{cons_law21}) and (\ref{cons_law22}) were derived first by York  in context of the energy-momentum tensor $T_{ab}$ in \cite{york} (see also \cite{eric}).}  We may replace $\mycal{G}_{ab}$, for instance, by $E_{ab}$. Accordingly, a simultaneous replacement of $\mathfrak{e}$, $\mathfrak{p}_a$ and $\mathfrak{S}_{ab}$ by $E^{{}^{(\mathcal{H})}}$, $\epsilon\,E^{{}^{(\mathcal{M})}}_a$ and $E^{{}^{(\mathcal{EVOL})}}_{ab}\hskip-.1cm{} + \kappa\, h_{ab} \,E^{{}^{(\mathcal{H})}}$, respectively, yields a system of equations which can be seen to be  equivalent to (\ref{constr2}) and (\ref{constr3}) whenever $E^{{}^{(\mathcal{EVOL})}}_{ab}\hskip-.1cm{}=0$. Note also that if the term  $E^{{}^{(\mathcal{EVOL})}}_{ab}$ is kept in these latter equations they can be used to justify the following statement which is complementary to that of Theorem\,
\ref{1stsym}.
\begin{lemma}
If the constraint expressions $E^{{}^{(\mathcal{H})}}$ and $E^{{}^{(\mathcal{M})}}_a$ vanish on all the $\sigma=const$ level surfaces then the relations 
\begin{eqnarray} 
K^{ab}\,E^{{}^{(\mathcal{EVOL})}}_{ab}\hskip-.1cm{}&\hskip-.2cm=&\hskip-.2cm{} 0\,, \label{cons_law210} \\ 
D^aE^{{}^{(\mathcal{EVOL})}}_{ab}\hskip-.1cm{} - \epsilon\,\dot n^a\,E^{{}^{(\mathcal{EVOL})}}_{ab}\hskip-.1cm{}  {}&\hskip-.2cm=&\hskip-.2cm{} 0\,. \label{cons_law220}
\end{eqnarray}
hold for the evolutionary expression $E^{{}^{(\mathcal{EVOL})}}_{ab}\hskip-.1cm{}$.
\end{lemma}
 
%%%%%%%%%%%%%%%%%%%% Consequences %%%%%%%%%%%%%%%%%%%%%%%%%%%%%%%%%%%
\section{Double decompositions}\label{integrability}
\setcounter{equation}{0}

Once a $1+n$ splitting has been done one may be interested in performing a succeeding $1+(n-1)$ decomposition provided that the $\sigma=const$ level surfaces are guaranteed to be foliated by a one-parameter family of $(n-1)$-dimensional hypersurfaces in $\Sigma_\sigma$. Note, however, that before automatically adopting Theorem \ref{1stsym} and the equations listed in the previous section the validity of all the assumptions made in deriving them have to be inspected. The key requirement to be checked is the vanishing of the covariant divergence of $\mycal{G}_{ab}$.
Therefore, once a $1+n$ decomposition had been done, before performing the succeeding $1+(n-1)$ splitting, we need to check whether the new source term, ${}^{{}^{(n)}}\hskip-1mm \mycal{G}_{ab}$, in  
\begin{equation}\label{geom2}
[{}^{{}^{(n)}}\hskip-1mm R_{ab}-\frac12\,h_{ab}{}^{{}^{(n)}}\hskip-1mm R\,]-{}^{{}^{(n)}}\hskip-1mm \mycal{G}_{ab}=0\,,
\end{equation}
does really have vanishing $D^a[{}^{{}^{(n)}}\hskip-1mm \mycal{G}_{ab}]$ divergence. In doing so notice first that 
\begin{equation}
{h^{e}}_{a}{h^{f}}_{b}\,[\, R_{ef}-\frac12\, g_{ef}\,R\,]={h^{e}}_{a}{h^{f}}_{b}\, R_{ef}-\frac12\, h_{ab}\,R
\end{equation}
and---by substituting (\ref{geom}) to the left hand side, whereas (\ref{expl_hhR}) and (\ref{curvature_relation}) to the right hand side---the source term can be seen to read as 
\begin{eqnarray}\label{int_evol_1+3} 
&& {}^{{}^{(n)}}\hskip-1mm \mycal{G}_{ab} = \mathfrak{S}_{ab} -\epsilon\left\{ -\mycal{L}_n K_{ab} - ({K^{e}}_{e}) K_{ab} + 2\,K_{ae}{K^{e}}_{b} - \frac{\epsilon}{N}\,D_a D_b N  \right. \\ && \left. \phantom{{}^{{}^{(n)}}\hskip-1mm \mycal{G}_{ab} = \mathfrak{S}_{ab}} + h_{ab}\left[\mycal{L}_n ({K^e}_{e}) + \frac12\,({K^{e}}_{e})^2   + \frac12\,K_{ef}{K^{ef}} +\frac{\epsilon}{N}\,D^e D_e N \right]\right\}  \nonumber \,. 
\end{eqnarray}

Notice that all the tensor fields involved in (\ref{int_evol_1+3}) are apparently fields defined on the $\Sigma_\sigma$ hypersurfaces thereby to proceed it suffices to ensure the existence of a foliation of $\Sigma_\sigma$ by a one parameter family of homologous codimension-two surfaces.

\medskip

Taking then the $D^a$-divergence of this relation and by commuting Lie and covariant, as well as, covariant derivatives, by a tedious but straightforward calculation, it can be verified that 
\begin{eqnarray}\label{div_evol_1+3} 
&& \hskip-1.0cm D^a  [{}^{{}^{(n)}}\hskip-1mm \mycal{G}_{ab}] = \mycal{L}_n\,\mathfrak{p}_b+ D^a \mathfrak{S}_{ab} +\epsilon\,\mycal{L}_n\,E^{{}^{(\mathcal{M})}}_b +\epsilon\,({K^e}_{e})\,\left[E^{{}^{(\mathcal{M})}}_b+\epsilon\,\mathfrak{p}_b\right] \nonumber  \\  
&& \phantom{ D^a {}^{{}^{(n)}}\hskip-1mm \mycal{G}_{ab} = }  \hskip-1.0cm + \dot n^a\,\left[ -\epsilon{}^{{}^{(n)}}\hskip-1mm R_{ab}+ \mycal{L}_n K_{ab}+ ({K^{e}}_{e}) K_{ab} - 2\,K_{ae}{K^{e}}_{b} + \frac{\epsilon}N\,D_a D_b N  
\right] \nonumber \\ &&  \phantom{ D^a {}^{{}^{(n)}}\hskip-1mm \mycal{G}_{ab} = }  \hskip-1.0cm - \dot n_b\,\left[ \mycal{L}_n ({K^e}_{e}) + K_{ef}{K^{ef}} + \frac{\epsilon}N\,D^e D_e N 
\right]  \,.
\end{eqnarray}

By inspecting (\ref{expl_hhR}) and (\ref{expl_nnR}), and the coefficients of $\dot n^a$ and $\dot n_b$ in (\ref{div_evol_1+3}) it can be recognized that they are equal to $-\epsilon\,{h^{e}}_{a}{h^{f}}_{b} R_{ef}$ and $n^e n^f R_{ef}$, respectively. Taking then into account  (\ref{geom}), along with $G_{ab}=R_{ab}-\frac12 g_{ab} R$, we get 
\begin{eqnarray}
{h^{e}}_{a}{h^{e}}_{a} R_{ef} {}&\hskip-.2cm=&\hskip-.2cm{} \mathfrak{S}_{ab}-\frac1{n-1}\,h_{ab}\,[\mathfrak{S}_{ef}\,h^{ef}+\epsilon\,\mathfrak{e} ] \\
n^e n^f R_{ef} {}&\hskip-.2cm=&\hskip-.2cm{} \mathfrak{e} -\frac{\epsilon}{n-1}\,[\mathfrak{S}_{ef}\,h^{ef}+\epsilon\,\mathfrak{e} ]\,.
\end{eqnarray} 
These relations, along with (\ref{div_evol_1+3}), imply that 
\begin{equation}\label{cons_law_rel}
D^a [{}^{{}^{(n)}}\hskip-.7mm \mycal{G}_{ab}]=0
\end{equation}
is equivalent to 
\begin{equation}\label{div_evol_1+3_e-1} 
\hskip-0.cm \mycal{L}_n\,\mathfrak{p}_b+ D^a \mathfrak{S}_{ab} + \left[-\epsilon\,\mathfrak{S}_{ab}\,\dot n^a +({K^e}_{e})\,\mathfrak{p}_b+\mathfrak{e}\,\dot n_b\,\right] + \epsilon\left[\mycal{L}_n\,E^{{}^{(\mathcal{M})}}_b + ({K^e}_{e})\,E^{{}^{(\mathcal{M})}}_b\right] =0 \,.
\end{equation}
In virtue of (\ref{cons_law22}) and (\ref{div_evol_1+3_e-1}) the integrability condition (\ref{cons_law_rel}) is guaranteed to hold whenever ${h^{f}}_{b}\,\nabla^a\,\mycal{G}_{af}=0$ and $E^{{}^{(\mathcal{M})}}_b=0$ on each of the $\sigma=const$ level surfaces. 

\medskip

In summarizing the above observations we have the following

\medskip

\begin{proposition}\label{cons_law_th}\vskip-0.2cm
The integrability condition (\ref{cons_law_rel}) holds on $\Sigma_\sigma$ if ${h^{f}}_{b}\,\nabla^a\,\mycal{G}_{af}$, the momentum constraint expression $E^{{}^{(\mathcal{M})}}_b$ and its Lie derivative $\mycal{L}_n\,E^{{}^{(\mathcal{M})}}_b$ vanish there.
\end{proposition}

In interpreting this result recall first that---by our assumptions concerning the source term for (\ref{geom})---the projection ${h^{f}}_{b}\,\nabla^a\,\mycal{G}_{af}$ vanish throughout $\Sigma_\sigma$. In addition, in virtue of Theorem \ref{1stsym} the Lie derivative of both the Hamiltonian and momentum constraint expressions vanish throughout $\Sigma_\sigma$ if they themselves vanish on $\Sigma_\sigma$ and the evolutionary system holds. Thus, as far as we prefer to solve first both the Hamiltonian and momentum constraints only on $\Sigma_\sigma$ we have to solve the reduced evolutionary system in $M$. In this case Proposition \ref{cons_law_th} has no use as it can guarantee the integrability condition for the reduced system after the solution has been found. 

\medskip

Note, however, that Proposition \ref{cons_law_th} allows a redistribution of the simplifications guaranteed by Theorem \ref{1stsym}. Namely, if we solve the momentum constraint on the entire base manifold in virtue of Theorem \ref{1stsym} and Proposition \ref{cons_law_th} besides solving the Hamiltonian constraint on $\Sigma_\sigma$ and instead of solving the full reduced system on $M$ it suffices to solve the second level of Hamiltonian and momentum constraints on a codimension-two surface in $M$ whereas the corresponding new reduced evolutionary system (formally only) on $\Sigma_\sigma$. By repeating this type of formal splittings\,\footnote{This could be done at most $n$-times which is the number of the equations involved in the original momentum constraint.} and solving always the yielded new momentum constraints the entire process can be applied inductively provided that product structure of the manifold allows it to be done. Applying this process, e.g.~to the conventional Cauchy problem in the 
Lorentzian case one may get on a suitable intermediate level a mixed elliptic-hyperbolic systems from Einstein's equations as it is done for a specific gauge choice in \cite{lars&vince}. 

%%%%%%%%%%%%%%%%%%%% FINAL REMARKS %%%%%%%%%%%%%%%%%%%%%%%%%%%%%%%%%%%
\section{Final remarks}\label{final}
\setcounter{equation}{0}

In answering the question raised in the title the main results of this paper makes it clear that some of the basic techniques developed for $1+n$ splitting of Lorentzian spacetimes do also apply to spaces with Riemannian metric. The most remarkable aspect here is that regardless whether the metric is of Euclidean or Lorentzian signature the subsidiary equations---these can be derived for the `Hamiltonian' and `momentum' type expressions, by making use of the Bianchi identity---are hyperbolic. This guaranties that in both cases the solutions to the `reduced' set of equations are also solutions to the full set of Einstein's equations (\ref{geom}) provided that the constraints hold on one of the hypersurfaces foliating the base manifold.
Having been the first $1+n$ type decomposition performed it is important to know if there may be room for further simplifications in a succeeding $1+(n-1)$ type decomposition. According to our findings there is no way to acquire new simplifications in a secondary splitting. 
It is remarkable that the new results apply regardless whether the primary space is Riemannian or Lorentzian.

\medskip

Having our results it would be useful to know whether they can be applied in solving some specific problems. To indicate that even in one of the simplest possible setup some non-trivial implications may follow let us consider the following example. Start with a four-dimensional Riemannian space foliated by a two-parameter family of homologous two-surfaces. Then---by making use of a suitable gauge fixing---the constraint equations can be seen to comprise a coupled parabolic-hyperbolic system, whereas the `evolutionary system' is comprised by coupled elliptic equations. The system corresponding to the constraint equations is under-determined and it has to be solved on one of the $\Sigma_\sigma$ hypersurfaces as a boundary-initial value problem with initial data specified on one of the codimension-two surfaces foliating $\Sigma_\sigma$. The other elliptic system corresponding to the `evolutionary' one in the present case has to be solved on the entire of the base manifold $M$ with boundary value yielded by the 
aforementioned parabolic-hyperbolic boundary-initial value problem on $\Sigma_\sigma$ (see e.g.~\cite{racz_geom_cauchy}). 

\medskip

The results covered by this paper have also applications in the conventional Cauchy problem and in the initial boundary value problem. In \cite{racz_geom_cauchy} it is demonstrated that the dynamics of four-dimensional spacetimes foliated by a two-parameter family of homologous two-surfaces can be interpreted as a two-surface based `geometrodynamics', whereas in \cite{racz_ibvp}---by making use of Proposition \ref{cons_law_th}, along with the fact that the results covered by Sections \ref{expl} and \ref{integrability} did not require any restriction on signature of the metric induced on the $\Sigma_\sigma$ hypersurfaces---some of the unsettled issues such as the geometric uniqueness in the metric based formulation of the initial boundary value problem will be addressed.

\medskip

It is worth emphasizing that concerning the metric only (\ref{geom}) had been used. This, besides the Riemannian or Lorentzian spaces satisfying Einstein's equations, allows many other theories, as well.\,\footnote{Note also that the $n$-dimensional source term in (\ref{int_evol_1+3}) cannot be directly connected to matter fields.} In particular, our assumptions are satisfied by the `conformally equivalent representation' of higher-curvature theories possessing a gravitational Lagrangian that is a polynomial of the Ricci scalar. Note also that the inclusion of metrics with Euclidean signature may significantly increase the variety of theories to be covered although no attempt has been made here to explore these aspects.

\medskip

Let us finally mention that irrespective of the simpleness of the observations made here they may have interesting applications elsewhere. It would be useful to know whether they could be used in string and brane theories, and also in various other alternative higher dimensional Riemannian and Lorentzian metric theories of gravity. 

%%%%%%%%%%%%%%%%%%%% ACKNOWLEDGMENTS %%%%%%%%%%%%%%%%%%%%%%%%%%%%%%%%%%%
\section*{Acknowledgments}

The author is grateful to Lars Anderson, Bobby Beig, G\'abor Zs.~T\'oth and Bob Wald for helpful comments and suggestions. 
This research was supported by the European Union and the State of Hungary, co-financed by the European Social Fund in the framework of T\'AMOP-4.2.4.A/2-11/1-2012-0001 ``National Excellence Program''. The author is also grateful to the Albert Einstein Institute in Golm, Germany for its kind hospitality where parts of the reported results were derived.

%%%%%%%%%%%%%%%%%%%% FROM HERE COME THE APPENDICES %%%%%%%%%%%%%%%%%%%%%
\appendix
\section{Appendix:}\label{Appendix A}
\renewcommand{\theequation}{A.\arabic{equation}}
\setcounter{equation}{0}

This section is to provide some useful relations. These had been applied in deriving several relations in Section \ref{expl}, and their adopted form will also be applied in our upcoming papers \cite{racz_geom_cauchy,racz_ibvp}. As our generic results are applicable in arbitrary dimension and to spaces with metric of Euclidean or Lorentzian signature we believe that these relations will find several applications.

\medskip   

It has been used implicitly in deriving (\ref{expl_nnR}) and it plays some role elsewhere so it is useful to give the generic relation of the scalar curvatures which reads as 
\begin{eqnarray}\label{curvature_relation}
&& R={}^{{}^{(n)}}\hskip-1mm R + \epsilon\,\left\{ -2\,\mycal{L}_{n} ({K^e}{}_{e}) - ({K^{e}}{}_{e})^2  - K_{ef} K^{ef} - \frac{2\,\epsilon}{N}\, D^e D_e N  \right\}\,.  
\end{eqnarray}

\medskip

Consider now a co-vector field $L_a$ on $M$ foliated by the $\sigma=const$ hypersurfaces. Then $L_a$ can be decomposed in terms of $n^a$ and fields living on the $\sigma=const$ level surfaces as
\begin{equation}
L_a={\delta^e}_a\,L_e=({h^e}_a+\epsilon\,n^en_a)\,L_e=\boldsymbol\lambda\, n_a + {\rm\bf L}{}_a\,
\end{equation}
where 
\begin{equation}
\boldsymbol\lambda= \epsilon\, n^e\,L_e \ \ \ {\rm and} \ \ \  {\rm\bf L}{}_a={h^e}_a\,L_e\,.
\end{equation}

Making use of this decomposition the covariant derivative $\nabla_e L_a$ and the divergence $\nabla^e L_e$ can be decomposed as 
\begin{eqnarray}
&& \hskip-0.8cm \nabla_e L_a= \left[\,D_e \boldsymbol\lambda + \epsilon\, n_e\,\mycal{L}_{n}\boldsymbol\lambda\,\right]\,n_a + \boldsymbol\lambda\,(K_{ea}+\epsilon\,n_e\dot n_a)+D_e {\rm\bf L}{}_a - n_e n_a \,(\dot n^f{\rm\bf L}{}_f) \\ && \hskip-0.8cm  \phantom{\nabla_e L_a= \left[(D_e \boldsymbol\lambda) + \epsilon\, n_e(\mycal{L}_{n}\boldsymbol\lambda)\right]\,n_a} + \epsilon\,\left\{n_e\,\mycal{L}_{n} {\rm\bf L}{}_a - n_e\,{\rm\bf L}{}_f {K^f}_a - n_a\,{\rm\bf L}{}_f {K^f}_e   \right\} \nonumber\,,
\end{eqnarray}
\begin{equation}
\nabla^e L_e=(h^{ea}+\epsilon\,n^en^a)\,\nabla_e L_a=\mycal{L}_{n}\boldsymbol\lambda+\boldsymbol\lambda\,({K^e}_e) + D^e {\rm\bf L}{}_e-\epsilon\,(\dot n^f{\rm\bf L}{}_f)\,.
\end{equation}

\medskip

Consider now a symmetric tensor $P_{ab}$ defined on $M$. Note first that $P_{ab}$ can be decomposed in terms of $n^a$ and fields living on the $\sigma=const$ level surfaces as
\begin{equation}
P_{ab}= \boldsymbol\pi \,n_a n_b  + \left[n_a \,{\rm\bf p}{}_b  + n_b\,{\rm\bf p}{}_a\right]  + {\rm\bf P}_{ab}\,,
\end{equation}
where $\boldsymbol\pi= n^e n^f\,P_{ef}$, ${\rm\bf p}{}_{a}=\epsilon\,{{h}^{e}}_{a} n^f\, P_{ef}$ and ${\rm\bf P}_{ab}={{h}^{e}}_{a}{{h}^{f}}_{b}\,P_{ef}$.

Then, the covariant derivative $\nabla_e P_{ab}$ can be decomposed as 
\begin{eqnarray}\label{Div_e_P_ab}
&& \hskip-0.8cm \nabla_e P_{ab} = \boldsymbol\pi\left[\,(K_{ea}+\epsilon\,n_e\dot n_a)\,n_b+(K_{eb}+\epsilon\,n_e\dot n_b)\,n_a\,\right] + \left[\,D_e \boldsymbol\pi + \epsilon\, n_e\,\mycal{L}_{n}\boldsymbol\pi\,\right]\,n_an_b \nonumber \\ 
&& \hskip-0.8cm \phantom{\nabla_e P_{ab} = } + (K_{ea}+\epsilon\,n_e\dot n_a)\,{\rm\bf p}{}_{b}+(K_{eb}+\epsilon\,n_e\dot n_b)\,{\rm\bf p}{}_{a} \nonumber\\ 
&& \hskip-0.8cm \phantom{\nabla_e P_{ab} = }  + n_a\left[\,D_e {\rm\bf p}{}_b+ \epsilon\,\left\{n_e\,\mycal{L}_{n} {\rm\bf p}{}_b - n_e\,{\rm\bf p}{}_f {K^f}_b - n_b\,{\rm\bf p}{}_f {K^f}_e   \right\} -n_e n_b \,(\dot n^f{\rm\bf p}{}_f) \,\right] \nonumber \\ 
&& \hskip-0.8cm \phantom{\nabla_e P_{ab} = }  + n_b\left[\,D_e {\rm\bf p}{}_a+ \epsilon\,\left\{n_e\,\mycal{L}_{n} {\rm\bf p}{}_a - n_e\,{\rm\bf p}{}_f {K^f}_a - n_a\,{\rm\bf p}{}_f {K^f}_e   \right\} -n_e n_a \,(\dot n^f{\rm\bf p}{}_f) \,\right] \nonumber \\ 
&& \hskip-0.8cm \phantom{\nabla_e P_{ab} = }  + \left[\,D_e {\rm\bf P}{}_{ab}+ \epsilon\,\left\{n_e\,[\,\mycal{L}_{n} {\rm\bf P}{}_{ab} - {\rm\bf P}{}_{fb} {K^f}_a - {\rm\bf P}{}_{af} {K^f}_b \,]   - n_a\,{\rm\bf P}{}_{fb} {K^f}_e - n_b\,{\rm\bf P}{}_{af} {K^f}_e \right\} \right. \nonumber \\ 
&& \left. \hskip-0.8cm \phantom{\nabla_e P_{ab} = + \left[D_e {\rm\bf p}{}_b\right. \ } -n_e n_b \,(\dot n^f{\rm\bf P}{}_{af}) -n_e n_a \,(\dot n^f{\rm\bf P}{}_{bf}) \,\right] \,,
\end{eqnarray}
while the contraction $\nabla^a P_{ab}$ reads as
\begin{eqnarray}\label{Div_P_ab}
&& \hskip-0.8cm \nabla^a P_{ab} = \boldsymbol\pi \,({K^e}_e)\,n_b+ ({K^e}_e) \,{\rm\bf p}{}_b + n_b\,(D^e {\rm\bf p}{}_e) + D^e {\rm\bf P}{}_{eb} - \epsilon\,n_b\,({\rm\bf P}{}_{ef}K^{ef})  \nonumber \\ 
&& \hskip-0.8cm \phantom{\nabla_e P_{ab} = } + \dot n_b\,\boldsymbol\pi + n_b\,\mycal{L}_{n}\boldsymbol\pi + \mycal{L}_{n} {\rm\bf p}{}_b - 2\,\epsilon\,(\dot n^e{\rm\bf p}{}_e)\,n_b - \epsilon\,(\dot n^e{\rm\bf P}{}_{eb})\,. 
\end{eqnarray}
The parallel and orthogonal parts of (\ref{Div_P_ab}) simplify as 
\begin{eqnarray}
\hskip-1.5cm (\nabla^a P_{ae})\, {{h}^{e}}_{b} {}&\hskip-.2cm=&\hskip-.2cm{} ({K^e}_e)\, {\rm\bf p}{}_b + D^e {\rm\bf P}{}_{eb} + \dot n_b\,\boldsymbol\pi + \mycal{L}_{n} {\rm\bf p}{}_b - \epsilon\,\dot n^e{\rm\bf P}{}_{eb} \,,\label{par_Div_P_ab} \\ 
\hskip-0.8cm (\nabla^a P_{ae})\,n^e {}&\hskip-.2cm=&\hskip-.2cm{} \epsilon\,[\,\boldsymbol\pi \,({K^e}_e) + D^e {\rm\bf p}{}_e - \epsilon\,{\rm\bf P}{}_{ef}K^{ef} + \mycal{L}_{n}\boldsymbol\pi   - 2\,\epsilon\,\dot n^e{\rm\bf p}{}_e\,]  \label{ort_Div_P_ab} \,. 
\end{eqnarray}

It also follows from (\ref{Div_e_P_ab}) that 
\begin{equation}
\nabla_a {P^e}_{e}= \epsilon\,\left[\,D_a \boldsymbol\pi + \epsilon\, n_a\,\mycal{L}_{n}\boldsymbol\pi\,\right] + D_a ({\rm\bf P}^e{}_{e}) +\epsilon\,n_a\,\mycal{L}_{n}({\rm\bf P}^e{}_{e})  \,,
\end{equation}
with parallel and orthogonal parts
\begin{eqnarray}
(\nabla_f {P^e}_{e})\,{{h}^{f}}_{a} {}&\hskip-.2cm=&\hskip-.2cm{} \epsilon\,D_a \boldsymbol\pi+D_a ({\rm\bf P}^e{}_{e}) \label{par_Div_P_ab_2} \\
(\nabla_f {P^e}_{e})\,{{n}^{f}} {}&\hskip-.2cm=&\hskip-.2cm{} \epsilon\,\mycal{L}_{n}\boldsymbol\pi + \mycal{L}_{n}({\rm\bf P}^e{}_{e})\,.\label{ort_Div_P_ab_2} 
\end{eqnarray}
%%%%%%%%%%%%%%%%%%%% REFERENCES %%%%%%%%%%%%%%%%%%%%%%%%%%%%%%%%%%%%%%%%
%\section*{References}

\end{document}